\documentclass[conference]{IEEEtran}

\usepackage{color,latexsym,amsfonts,amssymb}
\usepackage{amsmath,cite}
\usepackage{amsmath,amsthm,amsbsy}
\usepackage[mathscr]{euscript}
\usepackage[pdftex]{graphicx} 
\usepackage{tikz}
\usepackage{soul}
\usepackage[normalem]{ulem}

\usepackage{epstopdf} 

\usepackage{url}

\IEEEoverridecommandlockouts

\newcommand{\blue}[1]{\textcolor{black}{#1}} 

\newcommand{\stkout}[1]{\ifmmode\text{\sout{\ensuremath{#1}}}\else\sout{#1}\fi}

\newcommand{\E}{\mathbf{E}}

\newcommand{\Prob}{\mathbf{P}}

\newcommand{\R}{\mathbb{R}}

\newcommand{\calP}{\mathcal{P}}

\newcommand{\calK}{\mathcal{K}}

\newcommand{\Lap}{\mathcal{L}}
\newcommand{\SINR}{\textnormal{SINR}}

\newcommand{\threshold}{{\tau}} 
\newcommand{\noise}{{W}} 

\newcommand{\quality}{{q}}

\newcommand{\statespace}{{\cal{S}}}

\newcommand{\event}{\psi}

\newcommand{\kernelfun}{{\cal{K}}}

\newcommand{\psimod}{{\psi^{!i}}}

\theoremstyle{plain} 
\newtheorem{Theorem}{Theorem}[section]

\newtheorem{Lemma}[Theorem]{Lemma}
\newtheorem{IEEEproposition}[Theorem]{Proposition}

\theoremstyle{definition} 
\newtheorem{Example}{Example}[section]

\theoremstyle{remark} 
\newtheorem{Remark}{Remark}[section]

\begin{document}

\title{Coverage probability in  wireless networks with determinantal scheduling}

\author{B. B{\l}aszczyszyn, A. Brochard and H.P. Keeler
\thanks{{\bf B. B{\l}aszczyszyn} is with {\em Inria/ENS}, Paris, France; A. Brochard is with  {\em Inria} and {\em Huawei}, Paris, France; {\bf H.P. Keeler} is  with the {\em University of Melbourne} and {\em ACEMS}, Melbourne, Australia. This work was partly supported through Research Collaboration Agreement No. HF2016090005 between Huawei Technologies France and Inria
on {\em Mathematical Modeling of 5G Ultra Dense Wireless Networks}. H.P.Keeler was supported by an Australian Research Council DECRA  (ID: DE180100463).}
\vspace{-1ex}
}
\date{\today}
\maketitle

\begin{abstract}
We propose a new class of algorithms for randomly scheduling network transmissions. The idea is to use (discrete) determinantal point processes (subsets) to randomly assign medium access to  various {\em repulsive} subsets of potential transmitters. This approach can be seen as a  natural  extension of (spatial) Aloha, which schedules transmissions independently. Under a general path loss model and Rayleigh fading, we show that, similarly to Aloha, they are also subject to elegant analysis of the coverage probabilities and transmission attempts (also known as local delay). This is mainly due to the explicit, determinantal form of the conditional (Palm) distribution and closed-form  expressions for the Laplace functional of determinantal processes. Interestingly, the derived performance characteristics of the network are amenable to various optimizations of the scheduling parameters, which are determinantal kernels, allowing the use of techniques developed for statistical  learning with determinantal processes. Well-established sampling algorithms for determinantal processes can be used to cope with implementation issues, which is is beyond the scope of this paper, but it creates paths for further research.
\end{abstract}

\section{Introduction}
In wireless network research, an important challenge is scheduling the resources (essentially time, power and frequencies) to the network users, with the overall aim of allocating the service in some optimal manner. One  way to achieve such goals is to use (opportunistic) scheduling methods, where the algorithm makes decisions in real time, responding to the changes of the network and traffic demands. There is a large amount of research that proposes different scheduling techniques, which range from simple heuristics to complex mathematical algorithms ~\cite{asadi2013survey}. 

In general terms, schedulers are simply algorithms that wisely choose a subset of some underlying set with the aim of optimizing some utility function of that subset. Such subset selection problems are known to be, in general, very hard to solve computationally, as they are often NP-complete. One approach to overcome this difficult is to  leverage randomness. 

\subsection{Spatial Aloha}\label{ss.Aloha} 
For wireless networks, the problem of \emph{scheduling or subset selection} is usually dictated by \emph{medium access control} (MAC) protocols, with many being proposed over the years. A classic protocol based on probability is the (spatial) Aloha scheme in which network transmitters independently access the network with some probability $p$, where $p$ is a fixed constant sometimes called the \emph{medium access probability}. Proposed in the 1970s~\cite{abramson1970aloha}, this scheme has a long history and is particularly suitable for Poisson network models, where the transmitters are scattered across the plane $\R^2$ according to a Poisson point process $\Phi=\{X_i\}_i$ with intensity $\lambda>0$. Under the assumption of discrete time, at any time instant the transmitters accessing the network will form another Poisson point process  $\Phi^p=\{X_i\}_i$  with intensity $\lambda p$, due to the standard thinning result of the Poisson point process. 

Baccelli, B{\l}aszczyszyn and Singh~\cite{baccelli2014analysis} examined the case for transmitters being allowed to have different $p$ values depending on the network configuration $\Phi$. This ability to have different $p$ value for each transmitter motivated the term \emph{adaptive Aloha}.

\subsection{Determinantal point processes}
Researchers often build random models of wireless networks by using the Poisson point process to gain insight into the coverage probability of a single user based on its signal-to-interference-plus-noise ratio (SINR). But the Poisson point process does not exhibit repulsion between the points. To incorporate repulsion, researcher have developed network models using specific types of \emph{determinantal point processes}, obtaining coverage results~\cite{miyoshi2014cellular,nakata2014spatial,torrisi2014large,li2015statistical}.

Originally called \emph{fermion point processes}, determinantal  point processes  were first proposed to model repulsive particles, and they turn out to  have useful mathematical  properties~\cite{hough2006determinantal}, with recent research showing that they are particularly amenable to statistical inference methods~\cite{lavancier2015determinantal}. For wireless network models, these point processes are defined typically on the plane $\R^2$, but careful mathematical considerations and techniques are needed in this setting. These point processes are more tractable when they are defined on discrete spaces, meaning there is a finite number of possible point locations, often reducing mathematical technicalities to linear algebra. In this setting, Kulesza and Taskar~\cite{kulesza2012determinantal} used these point processes to develop a (supervised) machine or statistical learning framework for automatically choosing subsets, which uses the concepts of point quality and diversity; see Section~\ref{ss.simqual}, as well as other work by Kulesza and Taskar~\cite{kulesza2010structured,kulesza2012arxiv}. 

\subsection{Determinantal thinning}
\begin{figure}[t]
\begin{minipage}[b]{0.48\linewidth}
\centering
\centerline{\includegraphics[width=1.1\linewidth]{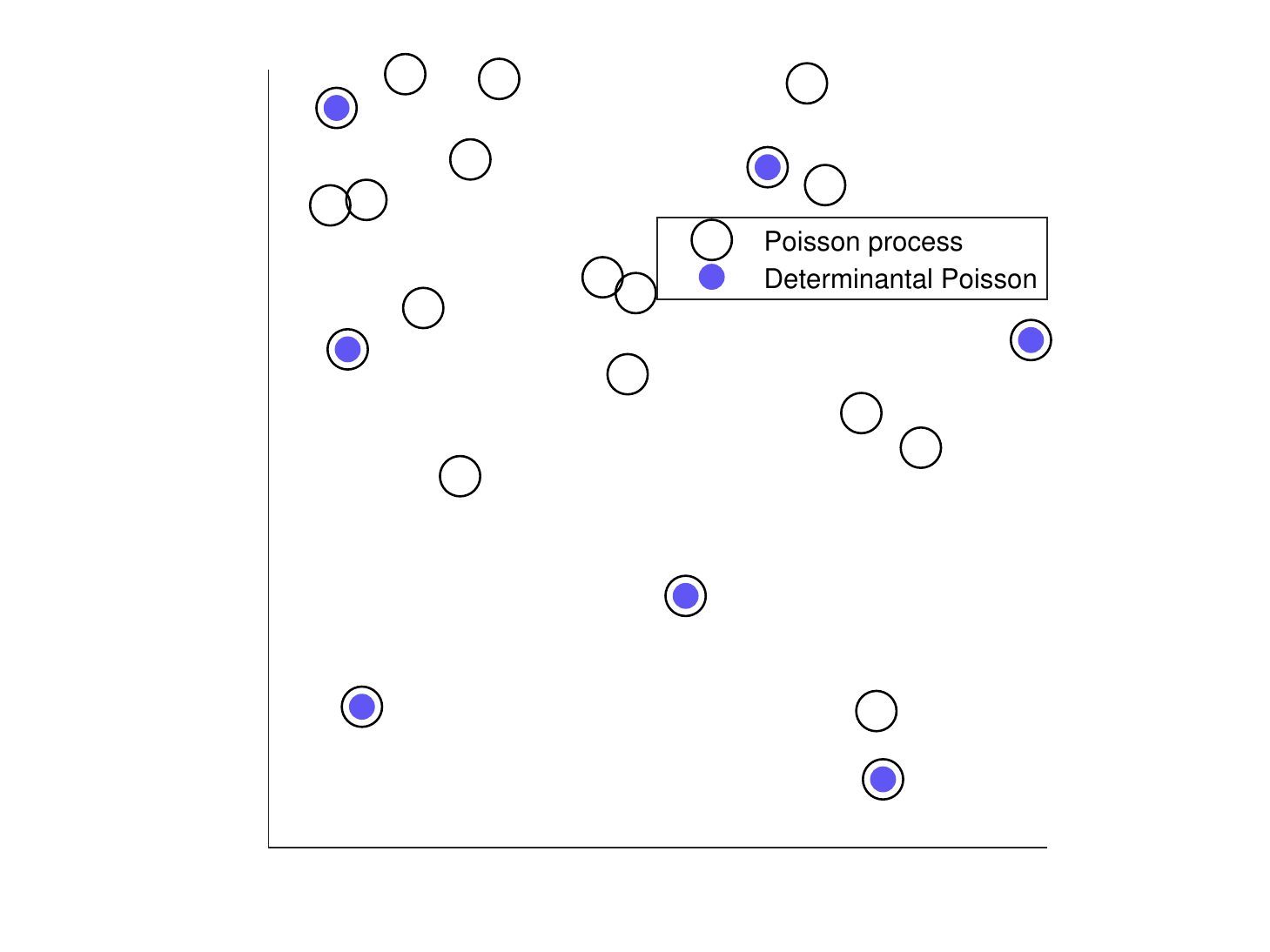}}
\vspace{-2ex}
\caption{\footnotesize A single realization of a determinantally-thinned Poisson point process, where the remaining points exhibit repulsion.}
\label{poissonthinned}
\end{minipage}
\hspace{0.1em}
\begin{minipage}[b]{0.48\linewidth}
\centering
\centerline{\includegraphics[width=1.1\linewidth]{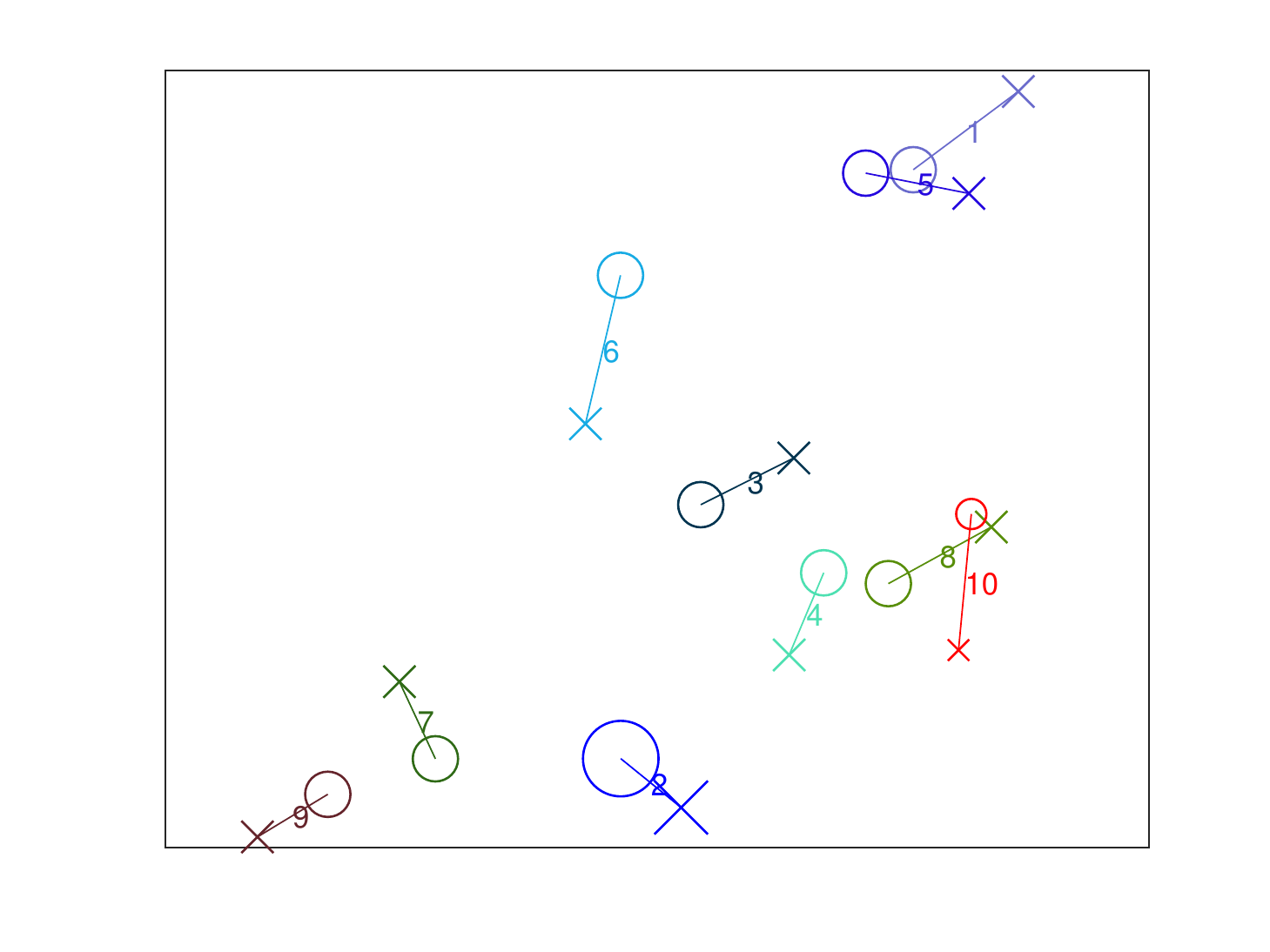}}
\vspace{-2ex}
\caption{\footnotesize A single configuration of transmitter-receiver pairs.}
\label{bipolar}
\end{minipage}
\vspace{-3ex}
\end{figure}

Recently B{\l}aszczyszyn and Keeler~\cite{blaszczyszyn2019determinantal} defined a new point process existing on bounded regions of the plane $\R^2$, which can be used as various types of network models; see Figure~\ref{poissonthinned}. They demonstrated that this new point process can be fitted to other repulsive point processes such as the Mat{\'e}rn hard-core types; also see~\cite{keeler2018detpoissoncode} for code and examples. This new point process is obtained by using a discrete determinantal point process to define a tractable point process operation called \emph{determinantal thinning}. Applied to the Poisson point process, B{\l}aszczyszyn and Keeler derived new results for the resulting point process, such as moment measures and Palm distributions, as well as an accompanying statistical fitting method based on maximum likelihoods. They also discussed the possibility of using determinantal point processes for scheduling wireless networks, with an emphasis on training the determinantal point processes on  pre-scheduled networks. This work initiated the current line of research.

\subsection{Other work}
Independently of our current work, Saha and Dhillon~\cite{saha2019machine} recently tackled the scheduling problem with discrete determinantal point processes. Theirs and our current approach only overlap slightly. They applied these point processes to the wireless link scheduling problem, with the aim of maximizing the overall network rate, by reducing the problem down to one of geometric programming, which is the subject of a recent review~\cite[Chapter 3 and Appendix A]{weeraddana2012weighted}. 

Conversely, our current work focuses on the special form of SINR problems, which allows for recasting the SINR expression into a tractable (matrix) kernel for determinantal point process. We achieve this by using the elegant algebraic properties of Palm distributions and Laplace functionals of determinantal points processes. 

\subsection{Determinantal extends Aloha}
Determinantal scheduling can be seen as a natural extension of the Aloha medium access scheme, where the former allows for repulsive patterns of simultaneous transmissions. In general, intuition says that the repulsive nature of determinantal point process should lead to better performance of the determinantal scheduler.

\subsection{Current contributions}
The current work contributes to the determinantal approach by treating the SINR coverage problem. Under the standard wireless signal propagation model, we show mathematically that the specific structure of the SINR problem combines elegantly with the special form of determinantal point processes, giving expressions for the coverage probabilities, which can then be immediately evaluated numerically. To reproduce the results, we have uploaded the code; see the accompanying repositories for the code in either MATLAB~\cite{keeler2020detcov_matlab} or Python~\cite{keeler2020detcov_python}.

We believe that the determinantal scheduling is particularly relevant as a medium access scheme for device-to-device networks, in which mobile users communicate directly with each other under some central authority of the covering base station, which is used to coordinate mutually dependent transmissions.  Well-established sampling procedures mean that operators can easily implement determinantal scheduling in wireless networks. We can also envision using a determinantal scheduling in other networks, such as a type of energy-saving sleep scheme~\cite{keeler2010stochastic,keeler2011model} or sentry selection~ \cite{balister2017sentry} in wireless sensor networks.

\section{Deterministic network model with random propagation effects}\label{s.network}
We now consider a popular physical model of a wireless network with deterministic transmitters and receivers, and  random propagation effects. We assume a deterministic (that is, non-random) configuration of potential transmitters on the plane written as $\phi=\{x_i\}_{i=1}^n\subset\R^2$\ .
 (The transmitters are on the plane, but the approach extends easily to higher dimensions.) For each transmitter $x_i\in \phi$, we consider  a receiver located at $y_i$, which in point process terminology is simply a \emph{mark}, resulting in a marked point process $\tilde{\phi}=\{(x_i,y_i)\}_{i=1}^n$, as illustrated in Figure~\ref{bipolar}. 

\subsection{SINR}
 We write $P_{x_i,y_i}$ to denote the received power at location $y_i$ of a signal emanating from a transmitter located at $x_i$. For a transmitter $x_i\in \phi$, consider its SINR at location $y_i\in\R^2$ with respect to the {\em active} (concurrent) transmissions from nodes in a configuration $\psimod\subset\phi$, to which transmitter $x_i$ does not belong, meaning $x_i\not\in\psimod$.  
 This SINR is  given by
\begin{equation}
\SINR(x_i,y_i; \psimod)= \frac{ P_{x_i,y_i}}{W+\sum\limits_{x_j\in \psimod} P_{x_j,y_i}  } \,,
\end{equation}
where $W$ denotes the noise power.

We now make some standard assumptions on the propagation model. For a signal traversing a distance $r$, we assume it undergoes path loss according to the deterministic, continuous, non-negative function $\ell(r)$, which is typically monotonic. A common choice is the power law
$\ell(r)=(\kappa r)^{-\beta }$,
where $\kappa>0$ and $\beta>0$, but our results hold for a more general path loss model, which is interesting in its own right. 

For any point $y\in \R^2 $, let {\em independent} random variable $F_{x_i,y_i}$ describe the random fading of the signal propagating from transmitter $x_i$ to location $y_i$. Then the SINR expression for transmitter $x_i\not\in\psimod$ becomes
\begin{equation}\label{e.SINR}
\SINR(x_i,y_i,\psimod)= \frac{ F_{x_i,y_i} \ell(x_i-y_i) }{\noise+\sum\limits_{x_j\in \psimod}  F_{x_j,y_i}\ell(x_j-y_i)  } \,.
\end{equation}

\subsection{Transmitter-and-receiver pairs}\label{s.covprobpp}
We consider a transmitter-and-receiver pair $(x_i,y_i)$ in the network $\tilde\phi$, so  $(x_i,y_i)\in\tilde\phi$, and a subset of active (interfering) transmitters $\psimod$ that does not include $x_i$. For $\threshold\geq0$, we write 
 \begin{equation}
 \Prob( \SINR(x_i,y_i,\psimod) > \threshold)    
 \end{equation} 
for the probability of a successful transmission from transmitter $x_i$ to its receiver $y_i$ given the set of active interferers $\psimod$. We call this probability the 
{\em pair-coverage probability between $x_i$ and $y_i$ against $\psimod$.}  This is the probability that the random propagation effects $F_{x_j,y_i}$ allow one to  achieve an SINR value in equation~\eqref{e.SINR} larger than some value $\threshold$. 

We now present a useful result for calculating these pair-coverage probabilities in a network with a path loss model $\ell$ and independent and identically-distributed (i.i.d.) Rayleigh fading. 
\begin{Lemma}[Elementary pair-coverage probabilities]\label{l.elem-covprob}
Assume  path loss model $\ell$ and i.i.d. exponential (with mean $\mu$) distribution of all $F_{x_j,y_i}$ variables. Then the pair-coverage probabilities between transmitter $x_i$ and its receiver $y_i$ against interferers $\psimod\not\owns x_i$ are given by
\begin{align}\label{e.elem-covprob}
\nonumber
\Prob&( \SINR(x_i,y_i,\psimod) > \threshold\,)=\\ 
&  w(|x_i-y_i|) \prod_{z \in \psimod} h(|z-y_i|,|x_i-y_i|)\,,
\end{align}
where the functions
\begin{align}\label{e.h2}
h(s,r)&:=\frac{1}{\threshold \ell(s)/\ell(r)+1}\quad s,r\ge0,\\
w(r)&:= \exp[- (\threshold/ \mu ) \noise / \ell(r)] 
\quad r\ge0.\label{e.w2}
\end{align} 
\end{Lemma}
We present the complete proof in Section~\ref{ss.proof-elem-covprob}, but we first note that Lemma~\ref{l.elem-covprob} is a variation of a result by B{\l}aszczyszyn and  M\"uhlethaler~\cite[Lemma 3.1]{linearNN} who proved it for the case of a power law path loss. Some additional factors appear in the original result due to the Aloha medium access scheme being assumed therein. 

To extend Lemma~\ref{l.elem-covprob}, we consider all the active transmitters $\psi:=\psimod\cup \{x_i\}$. The only source of randomness in our network model is the random Rayleigh fading.  But we can replace the non-random active set $\psi$ with a randomized subset $\Psi\subset\phi$ (resulting from a randomized scheduler or medium access scheme) conditioned on $x_i\in\Psi$. Then the product on the right-hand side of equation~\eqref{e.elem-covprob} takes the form of the probability generating functional of the \emph{reduced Palm} version of $\Psi$, a fact which we will use in the next section. 

\section{Coverage probabilities in a network with determinantal medium access}
\subsection{Determinantal medium access generalizes Aloha}\label{ss.DetMAC}
We define a \emph{determinantal (MAC) scheduler} as a random subset $\Psi\subset\phi$
whose distribution is a determinantal point process on the state space $\phi$
with some \emph{(marginal determinantal) kernel} $K$; see  Section~\ref{ss.det} for more details and  Section~\ref{ss.kernels} for kernel examples. For a real symmetric matrix $K$ indexed by the points of $\phi$, having all its eigenvalues in the interval $[0,1]$,
the distribution of $\Psi$ is characterized by the finite-dimensional probabilities
\begin{equation}\label{e.defdet}
\Prob(\Psi\supseteq  \event  ) = \det(K_{\psi}),    
\end{equation}
where $\det$ denotes the determinant, and $K_{\event}:=[K]_{x_i,x_j\in {\event}}$ denotes the restriction of $K$ to the scheduled points $\psi$.
 
\begin{Remark}
The determinantal (MAC) scheduler is a natural extension of the adaptive Aloha described in Section~\ref{ss.Aloha}, where we consider it here for a finite network. We can see this by taking the diagonal matrix kernel 
$K=\mathrm{diag}[(p(x_i))_{x_i\in\phi}]$,  for any function $p(x)$, which
makes $\Psi$ an independent Bernoulli thinning of $\phi$ with probabilities~$p(x_i)$.
\end{Remark}

We use a determinantal kernel $K$ on the set of transmitters $\phi$. We assume that the kernel $K=K(\tilde\phi)$ depends on the configuration of all potential transmitters and their receivers
$\tilde\phi=\{(x_i,y_i)\}$, but not on their fading variables.
As described in Section~\ref{ss.detthin}, $\Psi$ is a determinantal thinning of the transmitter configuration $\phi$.

To give the next results, we need some further notation. For a function $f(x_i)$, where $x_i\in \phi$, we write $K\{f\}$ to refer to a (matrix) kernel defined as
\begin{equation}
    [K\{f\}]_{x_i,x_j}:= \sqrt{1-f(x_i)}[K]_{x_i,x_j}\sqrt{1-f(x_j)}\,,
\end{equation}
where $x_i,x_j\in\phi$. Furthermore, for a  point $x_i\in\phi$, we define the  functions
\begin{equation}
h_{x_i}(x_j):=h(|x_j-y_i|,|x_i-y_i|),\quad x_j\in\phi\setminus\{x_i\}    \,,
\end{equation}
and constants 
\begin{equation}
W_{x_i,\blue{y_i}}:=w(|x_i-y_i|)\,,
\end{equation}
where functions $h$ and $w$ are given by expressions~\eqref{e.h2} and~\eqref{e.w2}.

Finally, Shirai and Takahashi~\cite[Theorem 1.7]{shirai2003random1} proved that the reduced Palm distribution of a determinantal point process is the distribution of another determinantal point process with a modified kernel; also see Section~\ref{ss.palm}. More precisely, 
for $x_i\in\phi$, we write  $K^!_{x_i}$ to denote the kernel indexed by points of $\phi\setminus\{x_i\}$ with entries
\begin{equation}\label{e.K-Palm}
[K_{x_i}^!]_{z_i,z_j}=[K]_{z_i,z_j}-
\frac{[K]_{z_i,x_i}[K]_{z_j,{x_i}}}{[K]_{{x_i},{x_i}}}\,,\quad z_i,z_j\in \phi\setminus\{{x_i}\}\,.
\end{equation}
In other words, the conditional distribution of $\Psi\setminus\{x_i\}$ given the point $x_i\in\Psi$ is also the distribution of another determinantal point process $\Psi_{x_i}^!$ with a modified kernel $K_{x_i}^!$ given by equation~\eqref{e.K-Palm}. \blue{The (non-reduced) Palm distribution is obtained by appending the point $x_i$ to the point process $\Psi_{x_i}^!$, which is characterized by the kernel $[K_{x_i}]_{z_i,z_j}$, where $z_i,z_j\in \phi$. For $ z_i,z_j\in  \phi\setminus\{{x_i}\}$, the kernel matrix entries coincide with that of $K_{x_i}^!$, meaning $[K_{x_i}]_{z_i,z_j}=[K_{x_i}^!]_{z_i,z_j}$, whereas the extra entries are one on the diagonal and zero elsewhere.  }

We now present our first main result for determinantal scheduler $\Psi$ on the configuration $\phi$ with  kernel~$K$. 
\begin{IEEEproposition}\label{p.SINR-conditional}
Assume a path loss model $\ell$ and i.i.d. exponential (with mean $\mu$) distribution of all fading variables. For a given transmitter $x_i\in\phi$ and SINR threshold $\threshold\ge0$, the SINR distribution at receiver $y_i$ from transmitter $x_i$, given $x_i$ is selected by the scheduler, is given by 
\begin{align}
\nonumber \Prob(\,&\SINR(x_i,y_i,\Psi\setminus\{x_i\})>\threshold\,|\,x_i\in\Psi\,) =
\\
&\det(I-K^!_{x_i}\{h_{x_i}\}) W_{x_i,\blue{y_i}} \, ,
\end{align}
where $I$ is an identity matrix. Furthermore, this probability also \emph{incorporates} the fading conditions and scheduler decisions.
\end{IEEEproposition}
The proof requires re-writing equation~\eqref{e.elem-covprob} in the form of a Laplace functional and a conditioning argument related to the reduced Palm distribution of~$\Psi$; see Section~\ref{ss.proof-SINR-conditional} for the complete proof.

For $\threshold\geq 0$, we denote by $\calP_i(\threshold)$ the unconditional SINR (tail) distribution at receiver $y_i$ from transmitter $x_i\in\phi$, 
\begin{equation}
\calP_i(\threshold):= \Prob(\,x_i\in\Psi\text{\ and\ \ } \SINR(x_i,y_i,\Psi\setminus \{x_i\})>\threshold\,)\,,
\end{equation}
which we call simply the \emph{coverage probability}. We present an expression for the coverage probability $\calP_i$.

\begin{IEEEproposition}\label{p.covprob}
The coverage probability is given by
\begin{equation}\label{e.covprob}
\calP_i(\threshold)=[K]_{x_i,x_i}\det(I-K^!_{x_i}\{h_{x_i}\})W_{x_i,\blue{y_i}}.
\end{equation}
\end{IEEEproposition}

\begin{IEEEproof}
Note that $\Prob\{x_i\in\Psi\}=
\det(K_{\{x_i\}})=[K]_{x_i,x_i}$. The result follows from Proposition~\ref{p.SINR-conditional} by conditioning on $x_i\in\Psi$.
\end{IEEEproof}

Note that the matrix $(I-K^!_{x_i})\{h_{x_i}\}$ on the right-hand-side of~\eqref{e.covprob} is indexed by the elements of $\phi\setminus\{x_i\}$. We extend it to the full configuration $\phi$
by taking
\begin{align}\label{e.Kt}    
\nonumber [&K_{x_i}(\threshold)]_{x_j,x_k} \\&:= \left\{
	\begin{array}{ll}
		[I - K_{x_i}^!\{h_{x_i}\}]_{x_j,x_k}  & \mbox{if } j,k\not=i \\
				W_{x_i,\blue{y_i}} [K]_{x_i,x_i}  & \mbox{if } j=k=i,\\
		0 &\mbox{if $j=i$ and $k\not=i$} \\
		&\mbox{or $k=i$ and $j\not=i$} \,.
			\end{array}
	\right. 
\end{align}
This new kernel allows us to express $\calP_i(\threshold)$ in a more compact way
\begin{equation}\label{e.covprob-det}
\calP_i(\threshold)=\det(K_{x_i}(\threshold)).     
\end{equation}

For a small number of pairs we can derive simple results.
\begin{Example}
Assume there are two transmitter pairs $\{x_1,y_1\}$ and $\{x_2,y_2\}$ with independent Rayleigh fading and a path loss model $\ell$. The  determinantal  kernel, which is always symmetric,  takes the general form 
\begin{equation}
K=\begin{bmatrix} 
k_{11} & k_{12} \\
k_{12} & k_{22} 
\end{bmatrix},
\end{equation}
 where the probabilities clearly $0\leq k_{11},k_{22} \leq 1$. (We need another condition so that that the eigenvalues of $K$ are also bounded on the unit interval.) 
 Then for the first transmitter-receiver pair, equation~\eqref{e.covprob-det} quickly gives the coverage probability 
\[
\calP_1= k_{11}[1-(k_{22}-k_{12}^2/k_{11})]\bar{h}_{12}w_1,
\]
where $    \bar{h}_{12}:=1- h_{x_1}(x_2)$ and $w_1:= w(|x_1-y_1|)$, which are constants in terms of optimizing the network scheduling. 
Symmetry gives the coverage probability for the other pair.
\end{Example}

\subsection{Transmission attempts (local delay)}
The quantity $\calP_i(\threshold)$ is the probability that a single transmitter $x_i$ can successfully transmit a message to its receiver $y_i$ in a single transmission attempt. But this may not occur in the first attempt, motivating us to examine the number of attempts needed for the first successful transmission, which Baccelli and B{\l}aszczyszyn~\cite{FBBBdelay09} called \emph{local delay}. 

For a transmitter $x_i\in\phi$, we will denote its local delay by $L_i=L(x_i,y_i,\phi)$. If we assume that at each attempt the determinantal scheduler $\Psi$ and fading variables $F_{x,y}$ are generated in an i.i.d. manner over successive time slots, then each $L_i$ has a geometric distribution.
\begin{Lemma}
 $L_i$ has a geometric distribution with the mean
\begin{equation}
  \E(L_i) =    \frac{1}{\calP_i(\threshold)} \, ,
\end{equation}
 which implies
\begin{align}
\Prob(L_i  \leq k) &= 1-[1-\calP_i(\threshold) ]^{k}  \,.    
\end{align}
\end{Lemma}
\begin{IEEEproof}
Given each attempt is independent of the previous one, $L_i$ is (by definition) a geometric variable with probability $\calP_i(\threshold)$ of success.
\end{IEEEproof}
The geometric distribution of $L_i$  allows for some straightforward analysis. For example, if a transmitter $x_i \in \phi$ makes $k$  transmission attempts, then for  the probability of having at least one successful transmission to be greater than some value $\epsilon>0$  the coverage probability needs to satisfy
\begin{equation}
    \calP_i(\threshold)  > 1- (1-\epsilon)^{1/k} \, .
\end{equation}

\section{Transmitter-or-receiver nodes}\label{s.TXorRX}
We now consider a single set $\phi$ of network nodes. We assume that if a node $x_i\in \phi$ is not transmitting, then it can be acting as a receiver. Under a determinantal scheduling scheme $\Psi$,  this means that a node $x_i \in \phi$ is transmitting if  $x_i\in \Psi$, and, conversely, a node $x_j\in \phi$ can  receive if $x_j \not\in \Psi$. We are interested in the probability of transmitting from node $x_i\in\phi$ to node $x_j\in\phi$, namely 
\begin{align}\label{e.prob-ij}
\nonumber\calP_{i,j}&(\threshold| x_i\in\Psi, x_j\not \in\Psi)\\
&:=\Prob(\SINR(x_i,x_j, \Psi\setminus\{x_i\})>\threshold | x_i\in\Psi, x_j\not \in\Psi)\,.
\end{align}

\blue{To present our next main result, we need the concept of a two-fold (reduced) Palm distribution for two points $x_i, x_j\in\Psi$, which we simply obtain by applying recursively expression~\eqref{e.K-Palm} to point $x_i$ then $x_j$ (or vice versa, as order does not matter). To be more specific, for a point $x\in \Psi$, we first write the reduced Palm distribution given by equation~\eqref{e.K-Palm} as $K_{x}^{!}:=\calK_{x}^!(K)$, where we interpret $\calK_{x}$ as a function (or an operator) working on the matrix $K$. Then the kernel for the two-fold reduced Palm distribution is 
\begin{equation}
    K_{x_i,x_j}^{!!}=\calK_{x_j}^!(\calK_{x_i}^!(K))\,,
\end{equation}
where we use the (non-standard) superscript notation ${!!}$ to highlight that this is the Palm distribution reduced by \emph{two} points. Similarly, the equivalent equation for the (non-reduced) Palm distribution is  $K_{x_i,x_j}:=\calK_{x_j}(\calK_{x_i}(K))$. }

\blue{In reality we will use a \emph{two-fold} Palm distribution reduced by only \emph{one point}, giving the ``semi-reduced'' Palm distribution kernel
\begin{equation}\label{e.semireduced}
    K_{x_i,x_j}^{!}=\calK_{x_j}(\calK_{x_i}^!(K))\,.
\end{equation}
We could use another approach, bypassing the need for this  kernel, but this is more suitable for practical implementations, as it allows an easy way to keep track of the indices of points and matrices when producing numerical results.}

We now present a result for calculating the  probability $\calP_{i,j}(\threshold| x_i\in\Psi, x_j\not \in\Psi)$; see Section~\ref{ss.proof-TXorRX} for the proof.
\begin{IEEEproposition}\label{p.TXorRX}
Assume path loss model $\ell$ and i.i.d. exponential (with mean $\mu$) distribution of all fading variables. In the network $\phi$ with determinantal scheduler $\Psi$, the probability of  transmitting from node $x_i$ to node $x_j$ is given by
\begin{align} \label{e.probTXorRX}
\calP_{i,j}(\threshold| x_i\in\Psi, x_j\not \in\Psi) &= \frac{W_{x_i,\blue{x_j}} }{(1-[K^{!}_{x_i}]_{x_j,x_j})} \\
\nonumber &\times \big[ \det(I-K^!_{x_i}\{h_{x_i}\}) \\
\nonumber&-[K_{x_i}]_{x_j,x_j} \det(I-K^!_{x_i,x_j}\{h_{x_i}\}) \big] \,,
\end{align}
where $W_{x_i,\blue{x_j}}=w(|x_i-x_j|)$,  function $w$ is defined by equation~\eqref{e.w2}, $K_{x_i}$ is a (non-reduced) Palm kernel, and $K^{!}_{x_i}$ and $K^!_{x_i,x_j}$ are respectively the reduced and semi-reduced (by $x_i$) Palm kernels.
\end{IEEEproposition}

\begin{Remark}
\blue{Provided a singular path loss model $\ell(r)=(\kappa r)^{-\beta }$,  if $x_j$ is a transmitter, then its signal power and, consequently, the interference are infinitely large at $x_j$ due to the singularity in the path loss model, resulting in $\SINR(x_i,x_j,, \Psi\setminus\{x_i\})=0$. Consequently, $\Prob(\SINR(x_i,x_j, \Psi\setminus\{x_i\})>\threshold | x_i\in\Psi, x_j  \in\Psi)=0$, which, in light of the  proof of Proposition~\ref{p.TXorRX}, reduces equation~\eqref{e.probTXorRX} to 
\begin{align} 
\nonumber\calP_{i,j}(\threshold| x_i\in\Psi, x_j\not \in\Psi) &=\frac{W_{x_i,\blue{x_j}}}{(1-[K^{!}_{x_i}]_{x_j,x_j}) } \\
&\times
 \big[ \det(I-K^!_{x_i}\{h_{x_i}\})  \big] \, ,
\end{align}
removing the need for the two-fold (semi-reduced) Palm distribution when $\ell(r)=(\kappa r)^{-\beta }$. This observation underscores the unexpected effects that a singular path loss model can have on SINR results. }
\end{Remark}

For $\threshold\geq0$, we denote by $\calP_{i,j}(\threshold)$ the unconditional SINR (tail) distribution at receiver $x_j$ from node $x_i\in\phi$, giving the coverage probability
\begin{align}
\nonumber\calP_{i,j}&(\threshold)\\
&:= \Prob(\,x_i\in\Psi, x_j\not \in\Psi\text{\ and\ \ } \SINR(x_i,x_j,\Psi\setminus \{x_i\})>\threshold\,)\,.
\end{align}
\begin{IEEEproposition}
The coverage probability is given by
\begin{equation}
\calP_{i,j}(\threshold):=\Prob( x_i\in\Psi, x_j\not \in\Psi)\calP_{i,j}(\threshold| x_i\in\Psi, x_j\not \in\Psi) \,.
\end{equation}
where 
\begin{equation}
\Prob( x_i\in\Psi, x_j\not \in\Psi)=[K]_{x_i,x_i}-\det(K_{\{x_i\}\cup \{x_j\}})\,.
\end{equation}
\end{IEEEproposition}
\begin{IEEEproof}
The proof is similar to that of Proposition~\ref{p.covprob}, in addition to $\Prob( x_i\in\Psi, x_j \in\Psi)=\det(K_{\{x_i\}\cup \{x_j\}})$, as given by expression~\eqref{e.defdet}.
\end{IEEEproof}

\section{Code}
We have implemented all our mathematical results into MATLAB and Python code, which is located in  the respective repositories~\cite{keeler2020detcov_matlab} and~\cite{keeler2020detcov_python}. We have also written the corresponding network simulations. The mathematical results agree excellently with simulations, which reminds us that determinantal point processes do not suffer from edge effects (induced by finite simulation windows). All mathematical and simulation results were obtained on a standard desktop machine, taking typically seconds to be executed. 

For a starting point, run the (self-contained) files \texttt{DemoDetPoisson.m} or \texttt{DemoDetPoisson.py} to simulate or sample a single determinantally-thinned Poisson point process. The determinantal simulation is also performed by the file \texttt{funSimSimpleDPP.m/py}, which requires the eigendecomposition of a $L$ kernel matrix; see Section~\ref{ss.Lensemble}.

The mathematical results for transmitter-and-receiver pair network, as described in Section~\ref{s.covprobpp}), are implemented in the file \texttt{ProbCovPairsDet.m/py}; also see \texttt{funProbCovPairsDet.m/py}. The  mathematical results for transmitter-or-receiver network, as described in Section~\ref{s.TXorRX}), are implemented in the file \texttt{ProbCovTXRXDet.m/py}; also see \texttt{funProbCovTXRXDet.m/py}. These files typically require other files located in the repositories. 

\section{Discussion and Conclusion}
In this work we have contributed to the line of research  that shows that determinantal point processes are particularly suitable as models of wireless networks. In addition to this, we have proposed the determinantal scheduler, which is amenable to analysis, particularly due to its mathematical properties in relation to the SINR. 

Our main new observation  is the mathematical form of the SINR problem (with a general path loss model and Rayleigh fading) allows one to treat SINR using results from determinantal point processes. Using this key insight, we derived new (linear algebraic)  expressions for the coverage probability (in terms of the SINR) for two wireless networks, in which we interpreted the new  determinantal scheduler as a MAC protocol. 

But of course the determinantal scheduler and our analysis can be applied to other types of wireless networks. For example, if one uses the determinantal scheduler as a sleep scheme in a wireless sensor network, then one could examine the probability of a certain region containing a sensor node that wakes up and successfully measures and relays information~\cite[Section 7]{keeler2011model}. 

In addition to the numerically tractable SINR expressions, the existence of efficient sampling algorithms for (discrete) determinantal point processes is yet another motivation to  use them as schedulers. They are also have good properties for statistically fitting data to models based them. 

Finally, practical research interest in determinantal point process (on discrete spaces) has  been strongly driven by the  (supervised) machine or statistical learning work pioneered by Kulesza and Taskar~\cite{kulesza2012determinantal} who showed that these point processes are very suitable for tackling subset selection problems. But this work has not been in  the context of wireless networks. We believe determinantal point processes show great promise as scheduling (or MAC) schemes. Moreover, we believe the tools of (determinantal) machine learning can be used to design schedulers and train them on pre-optimize wireless networks, opening up future research avenues. 

\appendix

\subsection{Proof of Proposition~\ref{l.elem-covprob}}\label{ss.proof-elem-covprob}
Given equation~\eqref{e.SINR}, we first consider the case of a single interferer $z = \psimod $ existing, which gives
\begin{align}
\nonumber \Prob&( \SINR(x_i,y_i,\psimod) > \threshold\,)\\ 
&=\Prob\{F_{(x_i,y_i)}\ell(|x_i-y_i|) \!\!\geq \!\!\threshold  (W +  F_{(z,y_i)}\ell(|y_i-z|)\}\\
  &=  \E[  e^{-(\threshold /\mu) W /\ell(|x_i-y_i|)  
  -(\threshold /\mu)F_{(z,y_i)} \ell(|z-y_i|) / \ell(|x_i-y_i|)}\bigr]
\\
  &=  e^{-(\threshold /\mu) W /\ell(|x_i-y_i|)} \E\bigl[e^{- (\threshold /\mu)  F_{(z,y_i)}
    \ell(|z-y_i|) / \ell(|x_i-y_i|)}\bigr]\,,
\end{align}
where we use the assumption that the $F$ variables are i.i.d. exponential random variables, and conditioning on $F_{(z,y_i)}$. (Recall the tail distribution $\Prob(E>t)=e^{-t/\mu}$ for an exponential random variable $E$ with mean $\mu$.) We see that the coefficient term gives the function $w$ defined by expression~\eqref{e.w2}.  Recalling the Laplace transform of an exponential variable $\E(e^{-E t})=1/(1+\mu t)$, the expectation term further reduces to
\begin{align}
\nonumber &\E\big[e^{- (\threshold /\mu) F_{(z,y_i)} \ell(|z-y_i|)/\ell(|x_i-y_i|)}\big]\\
&= [1+\threshold (\ell(|z-y_i|)/\ell(|x_i-y_i|)) ]^{-1}\,, 
\end{align}
which yields function $h$ defined by expression~\eqref{e.h2}. The rest of the proof follows from induction and independence of the $F$ variables.
$\blacksquare$

\subsection{Proof of Proposition~\ref{p.SINR-conditional}}\label{ss.proof-SINR-conditional}
Conditioning on the event $\Psi=\psimod\cup \{x_i\}$
and using~\eqref{e.elem-covprob}, we obtain 
\begin{align}\label{e.covprobpp}
\nonumber \Prob&( \SINR(x_i,y_i,\psimod) > \threshold \,| \Psi =\psimod\cup\{x_i\} \,)\\ 
\nonumber = & w(|x_i-y_i|) \\
&\times \exp\left(-\sum_{z \in \psimod} \log 1/h(|z-y_i|,|x_i-y_i|) \right) \,.
\end{align}
Observe that 
\begin{align}
 \nonumber \Prob&(\,\SINR(x_i,y_i,\Psi\setminus\{x_i\})>\threshold\,|\,x_i\in\Psi\,)\\
 &= \E_{x_i}^![  \Prob( \SINR(x_i,y_i,\psimod) > \threshold | \Psi=\psimod )] \,,
\end{align}
where the expectation $\E_{x_i}^!$ is taken with respect to the reduced Palm distribution of the (determinantal) point process $\Psi$. Then the expression takes the form 
\begin{align}
\nonumber \Prob&(\,\SINR(x_i,y_i,\Psi\setminus\{x_i\})>\threshold\,|\,x_i\in\Psi\,)\\
\nonumber= & w(|x_i-y_i|) \\
&\times \E_{x_i}^! \left[ \exp\left(-\sum_{z \in \Psi} \log 1/h(|z-y_i|,|x_i-y_i|) \right)\right] \,,
\end{align}
in which we see the Laplace functional of the point process $\Psi$ under $\E_{x_i}^!$, where we  denote this point process by $\Psi_{x_i}^!$, giving
\begin{align}
\nonumber \Prob&(\,\SINR(x_i,y_i,\Psi\setminus\{x_i\})>\threshold\,|\,x_i\in\Psi\,)\\
\nonumber &= w(|x_i-y_i|) \Lap_{\Psi_{x_i}^!}\left(-\log h(|z-y_i|,|x_i-y_i|) \right) \,.
\end{align}
The Laplace functional of a determinantal point process (detailed in Section~\ref{ss.laplace}) is expressed with a modified kernel given by equation~\eqref{e.Laplace}, yielding
\begin{equation}
     \Lap_{\Psi_{x_i}^!}\left(-\log h(|z-y_i|,|x_i-y_i|) \right) = \det[I-\bar{K}_{x_i}^!]\, ,
\end{equation}
where the kernel matrix $\bar{K}_{z}^!$, which is indexed by $z_i,z_j\in\phi\setminus\{x_i\}$, has the elements 
\begin{align}
\nonumber[\bar{K}_{x_i}^!]_{z_i,z_j}:=&[1-h(|z_i-y_i|,|x_i-y_i|)]^{1/2} \\
\nonumber\times &[K_{x_i}^!]_{z_i,z_j} \\
\times &[1-h(|z_j-y_i|,|x_i-y_i|)]^{1/2} \,.
\end{align}
The kernel of $\Psi_{x_i}^!$ is given by equation~\eqref{e.K-Palm}, completing the proof.
$\blacksquare$

\subsection{Proof of Proposition~\ref{p.TXorRX}}\label{ss.proof-TXorRX}
Define the events $A:=\SINR(x_i,x_j, \Psi\setminus\{x_i\})>\threshold\} $, $B:=\{x_i\in \Psi\}$, and $C:=\{x_j\not \in\Psi\}$, which implies $\Prob_{i,j}=\Prob(A|B,C)$, as defined by expression~\eqref{e.prob-ij}. Writing the complement of $C$ as $\bar{C}=x_j\in\Psi$, then the total law of probability gives
\begin{equation}\label{e.probAB}
    \Prob(A|B)= \Prob(C|B)\Prob(A|B,C)+\Prob(\bar{C}|B)\Prob(A|B,\bar{C}).
\end{equation}
We know that $1-\Prob(C|B)=\Prob(\bar{C}|B)=\Prob_{x_i}(x_j\in \psi)$, which is simply the diagonal element of (Palm) kernel $K_{x_i}$, so
\begin{equation}
    \Prob(\bar{C}|B)=[K_{x_i}]_{x_j,x_j} \,.
\end{equation}
The left-hand side of equation~\eqref{e.probAB} is 
\begin{equation}
    \Prob(A|B)= \Prob(\SINR(x_i,x_j, \Psi\setminus\{x_i\})>\threshold| x_i\in \Psi),
\end{equation}
which we obtain with Proposition~\ref{p.SINR-conditional}
by setting $x_j=y_i$. We also use Proposition~\ref{p.SINR-conditional} for the remaining probability term  
\begin{align}
\nonumber \Prob&(A|B,\bar{C})\\
&= \Prob(\SINR(x_i,x_j, \Psi\setminus\{x_i\})>\threshold| x_i,x_j\in \Psi)\\
 &= \E_{x_i,x_j}^![  \Prob( \SINR(x_i,x_j,\psimod) > \threshold | \Psi=\psimod )] \, ,
\end{align}
where the expectation $\E_{x_i,x_j}^!$ is taken with respect to the reduced (only by $x_i$) two-fold Palm distribution of the (determinantal) point process $\Psi$, as given by equation\eqref{e.semireduced}. In other words, this term is calculated by using the reduced (only by $x_i$) two-fold Palm distribution $\Prob_{x_i,x_j}^!$ conditioned on points $x_i$ and $x_j$ belonging to $\Psi$ (instead of using the one-fold Palm distribution). This distribution coincides with the distribution of a point process that is determinantal on $\phi\setminus (\{x_i\}\cup \{x_j\})$ with the kernel given by equation~\eqref{e.K-Palm} applied recursively to first $x_i$ and then $x_j$, which is then appended with  $x_j$, completing the proof. $\blacksquare$

\subsection{Determinantal point processes}\label{ss.det}
For a state space $\statespace$ with finite cardinality $m$, we consider a real symmetric $m\times m$ matrix  $K$ indexed by the points of $\statespace$, having  all its eigenvalues in the interval $[0,1]$. A discrete point process is a determinantal point process $\Psi$ defined with the kernel $K$ if for all configurations (or subsets) $\event \subseteq  \statespace $, the finite-dimensional probabilities are given by
\begin{equation}\label{e.dpp}
\Prob(\Psi\supseteq  \event  ) = \det(K_{\psi}) \,,
\end{equation}
where  $K_{\event}:=[K]_{x_i,x_j\in {\event}}$ denotes the restriction of $K$ to the entries indexed by the points in ${\event}$, that is $x_i, x_j\in \psi$.  Expression \eqref{e.dpp}~implies that each diagonal entry $K_{ii}$ of matrix $K$ is the probability of a location $x_i\in\statespace$ being occupied by a point of the point process $\Psi$, meaning 
$\Prob(x_i\in \Psi)=K_{ii} $. 
Expression \eqref{e.dpp} also means that a determinantal point process is not uniquely defined by a single matrix $K$.

\subsection{Kernel examples}\label{ss.kernels}
Usually one assumes the (marginal kernel) matrix $K$ to be positive semi-definite, recalling its  eigenvalues need to be bounded between zero and one. (One can also characterize $\Psi$ with a complex Hermitian kernel $K$.)
\begin{Example}\label{ex.kernel}
Consider a state space $\statespace$ consisting of three points, $\statespace=\{x_1,x_2,x_3\}\subset\R^2$, and define a determinantal point process with the kernel
\begin{equation}
K=\begin{bmatrix} 
k_{11} & k_{12} & k_{13} \\
k_{12} & k_{22}  & k_{23} \\ 
k_{13} & k_{23}& k_{33} 
\end{bmatrix},
\end{equation}
where the diagonals are probabilities, meaning $0\leq k_{11},k_{22},k_{33}\leq 1$. Then a further condition is needed to ensure that the eigenvalues of $K$ are also bounded on the unit interval.
\end{Example}

To populate the matrix $K$, one typically considers a kernel function $\cal{K}:\statespace\times\statespace \rightarrow \R $, such as the covariance functions used to define Gaussian processes. 
\begin{Example}
A typical example is the \emph{double exponential} or \emph{Gaussian kernel} function $$\kernelfun(x_i,x_j) =C e^{-(|x_i-x_j|^2/\sigma^2)},$$ where $x_i,x_j \in \statespace$, $\sigma>0 $, and $C>0$ is some suitable constant that ensures the eigenvalues of $K$ are properly bounded. 
\end{Example}
It is not always obvious how to define new kernels, particularly when needing properly bounded eigenvalues, but we will see in the Section~\ref{ss.Lensemble} that they can be easily defined by using the formalism of $L$-ensembles.

\subsection{Kernels using $L$-ensembles}\label{ss.Lensemble}
The eigenvalues of the (non-negative definite) matrix $L$ are real, non-negative, but need not be smaller than one. $K$ has the same the eigenvectors as $L$, but its eigenvalues are equal to 
$\lambda_i/(1+\lambda_i)$, where $\lambda_i$ are the eigenvalues of $L$. Any determinantal kernel $K$ having all eigenvalues strictly smaller than one has the form~\eqref{e.K-L} with $L=(I-K)^{-1}-I$. Moreover, we have for all $\psi\subset\statespace$ the probabilities
\begin{equation}\label{e.L}
\Prob\{\,\Psi=\event\,\} = \frac{\det(L_{\event}) }{\det{(L+I)}}.
\end{equation} 
These types of determinantal point processes are called {\em $L$-ensembles}. They were originally studied in mathematical physics~\cite{borodin2005eynard}, but Kulesza and Taskar~\cite{kulesza2012determinantal} applied them to the machine learning problem of subset selection, building off their tractability for defining suitable kernels. 

\subsection{Similarity and quality}\label{ss.simqual}
A special class of determinantal point processes have  kernels of the form
\begin{equation}\label{e.K-L}
K=L(L+I)^{-1}
\end{equation}
for some real, symmetric, non-negative definite 
matrix~$L$; see Section~\ref{ss.Lensemble} for more details. We briefly recall the approach proposed by Kulesza and Taskar~\cite{kulesza2012determinantal}. Consider a matrix $L$ whose elements can be written as
\begin{equation}\label{e.Ldecomp}
[L]_{x_i,x_j} = \quality_{x_i} \,[S]_{x_i,x_j}\, \quality_{x_j} ,
\end{equation}
for $x_i,x_j\in\phi$, where  $q_{x}$ is a positive function of $x\in\phi$ and $S$ is a symmetric, positive semi-definite $m\times m$ matrix, where $m=\#(\phi)$. These two terms are known as \emph{quality} and the \emph{similarity matrix}. The quality $q_x$ measures the goodness of point $q_x\in\phi$, while $[S]_{x_i,x_j}$ gives a measure of similarity between points $x_i$ and $x_j$.  The larger the $q_x$ value, the more likely there will be a point of the determinantal point process at location~$x$, while the larger  $[S]_{x_i,x_j}$ value for two locations $x_i$ and $x_j$, the less likely realizations will occur with two points simultaneously at both locations.

\subsection{Palm distributions}\label{ss.palm}
Determinantal point processes exhibit closure under Palm distributions. Shirai and Takahasi~\cite[Theorem 1.7]{shirai2003random1} proved that for a general determinantal point processes, its Palm distribution coincides with the probability distribution of another determinantal point process with a modified kernel. 

\begin{Example}\label{ex.Palm}
We recall the kernel of the three-point state space given in Example~\ref{ex.kernel}.
For  point $x_1$, the reduced Palm version of the kernel is
\begin{equation}
K_{x_1}^!=    \begin{bmatrix} 
 k_{22}  & k_{23} \\ 
k_{23}  &  k_{33}
\end{bmatrix} -\frac{1}{k_{11}}
\begin{bmatrix} 
 k_{12}^2 & k_{12}k_{13} \\ 
k_{12}k_{13}  &  k_{13}^2
\end{bmatrix} 
\end{equation}
\end{Example}

Shirai and Takahasi~\cite[Corollary 6.6]{shirai2003random1} obtained similar Palm results for the $n$-fold Palm distribution when conditioning on multiple points. Borodin and Rains~\cite[Proposition 1.2]{borodin2005eynard} independently derived Palm distributions for $L$-ensembles. But the connection between these two results has only been investigated recently~\cite[Section 5.7.4]{baccelli2020random}. B{\l}aszczyszyn and Keeler~\cite[Section B]{blaszczyszyn2019determinantal} presented another proof of the Borodin and Rains result.

\subsection{Laplace functional}\label{ss.laplace}
For any non-negative function $f$, the Laplace functional of the detetermintal point process $\Psi$ is given by
\begin{align}\nonumber
\textbf{L}_{\Psi}(f):&=\E\,\left[e^{-\sum_{x\in \Psi} f(x) } \right] \\
&=\det[I-K']\, ,\label{e.Laplace}
\end{align}
where the kernel matrix $K'$ has the elements 
\begin{equation}\label{e.Kernel-Laplace}
[K']_{z_i,z_j}:=[1-e^{-f(z_i)}]^{1/2}\, [K]_{z_i,z_j}\, [1-e^{-f(z_j)}]^{1/2},
\end{equation}
for all $z_i,z_j\in\phi$.  Shirai and Takahashi~\cite{shirai2003random2} proved this in the general discrete case. But B{\l}aszczyszyn and Keeler~\cite[Section A]{blaszczyszyn2019determinantal} presented a simpler, probabilistic proof, which uses the finite state space assumption of the determinantal process.

\subsection{Determinantal thinning}\label{ss.detthin}
To define determinantal thinning, we consider a (non-random) point pattern $\phi$ on some bounded region ${\cal{R}}\subset \R^2 $. For the determinantal point process $\Psi$, we set the state space as the point pattern, so $\statespace=\phi$, resulting in a thinned point pattern.  More precisely, the points of the point pattern $\phi$ form the state space of the finite determinantal point processes. The point process  {$\Psi $} is defined on a subset of the plane {$\R^2$}, and the points of the point pattern {$\phi$} are dependently thinned such that there is repulsion among the points of  {$\Psi$}.  

\begin{Example}
The point pattern $\Phi$ is a single realization of a homogeneous Poisson point process $\Phi$ with intensity $\lambda>0$ defined on a bounded region ${\cal{R}}\subset \R^2 $, which implies the conditional probabilities
\begin{equation}\label{e.dppp}
\Prob(\Psi\supseteq  \event |\Phi=\phi ) = \det(K_{\psi})\,,
\end{equation}
where $K_{\event}=[K]_{x_i,x_j\in {\event}}$ such that all $x_i,x_j\in \phi\subset {\cal{R}} $. This conditioning argument leads to the determinantally-thinned Poisson point process introduced by Keeler and B{\l}aszczyszyn~\cite{blaszczyszyn2019determinantal}, which is illustrated in Figure~\ref{poissonthinned}.
\end{Example}

\addtocounter{section}{1}
\addcontentsline{toc}{section}{References}

\end{document}